\documentclass[twocolumn,preprintnumbers,amsmath,amssymb]{revtex4}

\usepackage{graphicx}

\bibpunct{[}{]}{,}{n}{,}{,}

\begin{document}

\preprint{Preprint Dated \today}

\title{\large Scaling of exciton binding energy with external dielectric function in carbon nanotubes}

\author{Andrew G. Walsh$^{a}$}
\author{A. Nickolas Vamivakas$^{b}$}
\author{Yan Yin$^{a}$}
\author{Stephen B. Cronin$^{c}$}
\author{M. Selim \"{U}nl\"{u}$^{b,a}$}
\author{\\Bennett B. Goldberg$^{a,b}$}
\author{Anna K. Swan$^{b,a,}$\footnote{Corresponding author.\\\textit{E-mail address:} swan@bu.edu.}}
\affiliation{\fontfamily{ptm}\selectfont $^{a}$Department of Physics, Boston University, 590 Commonwealth Avenue, Boston, MA 02215\\$^{b}$Department of Electrical and Computer Engineering, Boston University, 8 Saint Mary's Street, Boston, MA 02215\\$^{c}$Department of Electrical Engineering, University of Southern California, Powell Hall of Engineering PHE 624, Los Angeles, CA 90089-0271}

\begin{abstract}
\begin{flushleft}

\line(1,0){500}

\noindent
\\[5pt]
{\bf Abstract}\\[8pt]

~~We develop a scaling relationship between the exciton binding energy and the external dielectric function in carbon nanotubes.  We show that the electron-electron and electron-hole interaction energies are strongly affected by screening yet largely counteract each other, resulting in much smaller changes in the optical transition energy.  The model indicates that the relevant particle interaction energies are reduced by as much as 50 percent upon screening by water and that the unscreened electron-electron interaction energy is larger than the unscreened electron-hole interaction energy, in agreement with explanations of the ``ratio problem.''  We apply the model to measurements of the changes in the optical transistion energies in single, suspended carbon nanotubes as the external dielectric environment is altered.\\

\noindent

\copyright \ 2007 Elsevier B. V. All rights reserved.\\[8pt]



\line(1,0){500}

\end{flushleft}
\end{abstract}

\maketitle

\fontfamily{ptm}\selectfont

{\bf 1. Introduction}\\

The optical and electronic properties of single wall carbon nanotubes (SWCNTs) are known to be dominated by strong Coulomb interactions between electrons and electrons and between electrons and holes. \cite{Ando,Kane1,Kane2,Spataru,Perebeinos} Two-photon experiments have measured exciton binding energies of several hundred meV. \cite{Wang2,Maultzsch1,Dukovic}  However, these measurements were performed on SWCNTs in screened environments; the intrinsic, chirality dependent, unscreened exciton binding energies can be significantly larger\cite{Capaz3}.  Understanding how these particle interaction energies change with screening by the nanotube environment is critical when designing opto-electronic devices, carbon nanotube field effect transistors, etc.  Previous theoretical models of particle interaction energies in carbon nanotubes (CNTs) typically include a single variable for the dielectric function and treat it as a fit parameter.  Thus, in these models, the dielectric function represents some average value of the heterogeneous dielectric environment and can not be used as an input parameter even when the value of the dieletric function external to the CNT is known.  In this paper, we derive a scaling relationship the uses the actual external dielectric function.  We use this model in Ref. \cite{Walsh} to fit resonance Raman data taken from single CNTs suspended across trenches as the dielectric environment is altered.  The results show that the particle interaction energies are about two times larger in air than when screened in water.  However the measured energy shift of the optical transition energy is small since the band gap renormalization and exciton binding energies have opposite signs. \cite{Ando,Kane1}  That is, the changes in these underlying interaction energies may be separately quite large but their difference relatively small, in agreement with reported solvatochromic shifts\cite{Choi,Lefebvre1,Moore}.\\

{\bf 2. Theory}\\

\begin{figure}
\begin{center}
\includegraphics[width=3.0in]{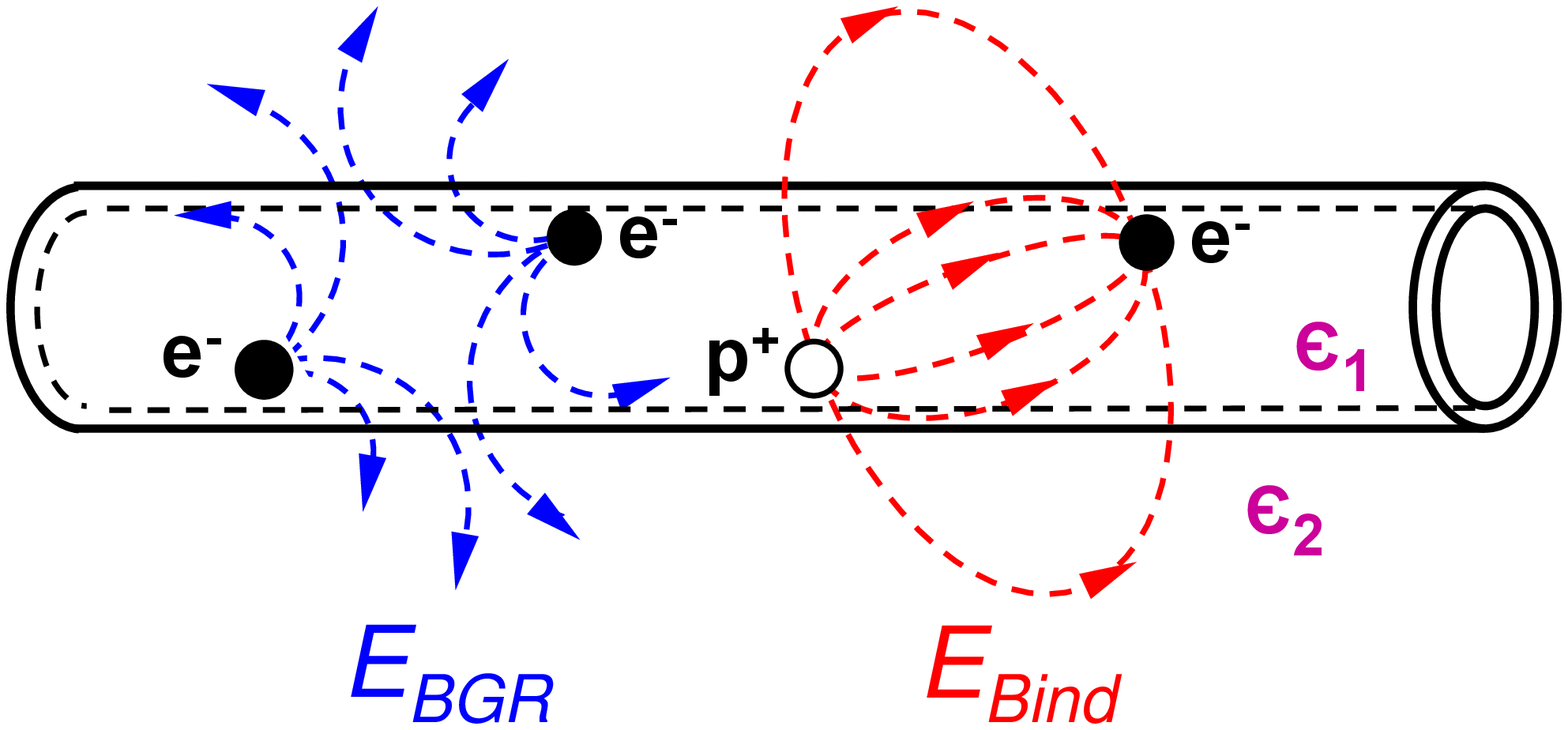}
\caption{Screening of the field lines between electrons and holes in a carbon nanotube of dielectric value $\epsilon_{1}$ in a dielectric environment of $\epsilon_{2}$.  Electron-electron and electron-hole interactions lead to \textit{E}$_{BGR}$ and \textit{E}$_{Bind}$, respectively.}
\label{fig1}
\end{center}
\end{figure}

The one-dimensional nature of carbon nanotubes leads to smaller Coulomb screening and larger particle interactions compared to two- and three-dimensional materials.  Thus, Coulomb interaction energies can not be ignored when attempting to understand the electronic structure in one-dimensional systems.  The one-dimensional nature of carbon nanotubes also makes their electronic structure very sensitive to their environment and changes therein.  Fig. \ref{fig1} schematically depicts the screening of (or lack thereof) the electric field lines in carbon nanotubes of dielectric value $\epsilon_{1}$ in an external dielectric environment $\epsilon_{2}$.  Electronic interactions lead to large blue shifts of the free particle band gap, a process known as band gap renormalization, while electron-hole interactions lead to a series of bound excitonic states well inside the band gap. \cite{Ando}  We label these interaction energies \textit{E}$_{BGR}$ and \textit{E}$_{Bind}$ (refering to the lowest optically active exciton which dominates the optical response\cite{Vamivakas}), respectively.  In fact, these particle interaction energies in an unscreened environment are calculated to be on the order of one electron volt for one nanometer diameter single wall carbon nanotubes, about the size of the single particle band gap.  However, since they enter into the Hamiltonian with opposite signs,\cite{Kane1} they largely cancel each other and the resulting optical transition energy, \textit{E}$_{Opt}$, is only slightly higher than the transition energy predicted by single particle models, \textit{E}$_{SP}$.  That is, \textit{E}$_{Opt}$ = \textit{E}$_{SP}$ + \textit{E}$_{BGR}$ + \textit{E}$_{Bind}$.  We now derive an expression for the scaling dependence of the exciton binding energy, \textit{E}$_{Bind}$, on the external dielectric, $\epsilon_{2}$.

\begin{figure}
\begin{center}
\includegraphics[width=3.3in]{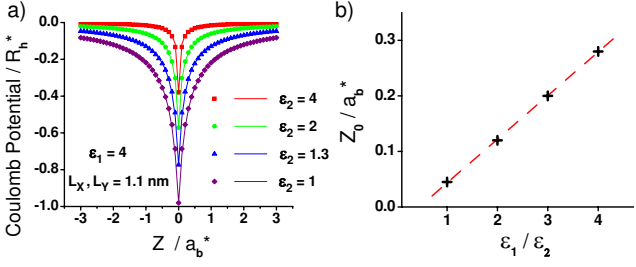}
\caption{(a) V$_{1D}^{Eff}$ as a function of electron hole separation, z, for four values of the external dielectric $\epsilon_{2}$.  Truncated Coulomb potential fits to the numerically calculated data are shown as solid lines.  a$_{b}^{*}$=$\epsilon_{1}\hbar^{2}$/$\mu$e$^{2}$ is the bulk exciton Bohr radius.  (b) The best fit cut off parameter z$_{0}$ from (a) as a function of external dielectric $\epsilon_{2}$.  The dashed red line highlights the linear dependence of z$_{0}$ on $\epsilon_{2}$.}
\label{fig2}
\end{center}
\end{figure}

We use a one-dimensional effective potential, V$_{1D}^{Eff}$(z), for a quantum wire of dielectric $\epsilon_{1}$ in an environment $\epsilon_{2}$, integrating over the lateral x, y dimensions which yields a function of z, the electron-hole separation. \cite{Ogawa1,Banyai}  Numerical results from Ref. \cite{Ogawa1} are shown in Fig. \ref{fig2}a for four different values of the ratio of $\epsilon_{1} / \epsilon_{2}$.  $\epsilon_{1}$ is taken as 4 for graphite. \cite{Pedersen2, Taft} $\epsilon_{2}$, the external dielectric, spans values from 1, i.e. unscreened, to 4.  Here, the binding energy of the exciton dictates that we use the optical value of the dielectric function, which, for water, is about 1.78\cite{Hayashi}.  The resulting curves are fit with a truncated Coulomb potential of the form 1 / $\vert$z$\vert$+z$_{0}$.  The fit parameter z$_{0}$ is known as the cutoff parameter.   By incorporating z$_{0}$, the divergence as the electron-hole separation z$\rightarrow$ 0 is removed.  This also reflects the geometry of the problem, i.e. the carbon nanotube is not truly one-dimensional but has a finite diameter implying a minimum electron-hole separation.  The fits are clearly excellent.  The best fit value of z$_{0}$ for each ratio $\epsilon_{1} / \epsilon_{2}$ is plotted in Fig. \ref{fig2}b.  For the range of external dielectric values of interest here, the dependence is almost perfectly linear.  That is, we can say z$_{0}$ scales with $\epsilon_{2}^{-1}$.  The same scaling relationship is found using the expression for the one-dimensional effective potential of Ref. \cite{Banyai} as well.

This result allows us to derive a scaling relationship between the exciton binding energy and the external dielectric function using Ref. \cite{Loudon} which analytically solved the binding energy for the one dimensional hydrogen atom using the truncated Coulomb potential.  Specifically, Ref. \cite{Loudon} found that the binding energy \textit{E}$_{Bind}$=R$^{*}$ / $\lambda^{2}$ where R$^{*}$ is an effective Rydberg equal to $\mu$e$^{4}$ / 2$\hbar^{2}\epsilon^{2}$, $\mu$ is the exciton effective mass, and $\epsilon$ is the dielectric constant, a poorly defined quantity when applied to a heterogeneous environment.  The quantum number, $\lambda$, is not necessarily integer and is a complicated function of the cutoff parameter, z$_{0}$.  However, Ref. \cite{Combescot} shows that $\lambda$ scales as $\sim$ z$_{0}^{0.4}$ which, in turn, we have shown scales as $\epsilon_{2}^{-1}$.  Thus, \textit{E}$_{Bind}$ scales as R$^{*}$ x $\epsilon^{2*0.4}$.  Further, the depth of the effective potential, which is proportional to e$^{2}$ / $\epsilon$, on Fig. \ref{fig2}a is found to scale with 1 / $\epsilon_{2}$ over this external dielectric range.  Thus, the effective Rydberg, R$^{*}$, which is proportional to e$^{4}$ / $\epsilon^{2}$, is presumed to scale as 1 / $\epsilon_{2}^{2}$.  The overall scaling of the exciton binding energy is then $\epsilon^{2*0.4}$ / $\epsilon^{2}$ = 1 / $\epsilon^{1.2}$, the central result of this paper.  We emphasize that this scaling relationship is based on the actual external dielectric value and is thus of practical use.  The value of the scaling exponent, $\alpha$, is very close to the value $\alpha$ = 1.4 derived in Ref. \cite{Perebeinos} where the model contained a single $\epsilon$ and thus represented a sort of averaging over the heterogeneous dielectric environment.  Also, since R$^{*}$ is independent of the radius, r , and z$_{0}$ scales with r,\cite{Ogawa1,Ogawa2} the bindng energy scales with 1 / r$^{0.8}$, close to the 1 / r$^{0.6}$ dependence found by Ref. \cite{Pedersen1} using a variational method.

We now combine this scaling result with the scaling behavior of the bad gap renormalization energy, \textit{E}$_{BGR}$, in order to address how the eletronic structure, which depends on both electron-electron and electron-hole interactions, scales with the external dielectric environment.  Specifically, Ref. \cite{Kane1} finds \textit{E}$_{BGR}$ scales approximately as a 1 / $\epsilon$.   Having found the dependence of R$_{*}$ on $\epsilon_{2}$ was the same as the dependence on $\epsilon$ for this dielectric range, we assume \textit{E}$_{BGR}$ scales with 1 / $\epsilon_{2}$.  Then the expression \textit{E}$_{Opt}$ = \textit{E}$_{SP}$ + \textit{E}$_{BGR}$ + \textit{E}$_{Bind}$ becomes\\

\textit{E}$_{Opt}(\epsilon_{2})$ = \textit{E}$_{SP}$ + \textit{E}$_{BGR}^{\epsilon_{2}=1}$/$\epsilon_{2}$ + \textit{E}$_{Bind}^{\epsilon_{2}=1}$/$\epsilon_{2}^{1.2}$,\\

\noindent ignoring the possible small dependence of the single particle term on external dielectric through the exciton effective mass, $\mu$, which is a function of valence and conduction band curvature.  Thus, upon changing the environmental screening,\\

$\Delta$\textit{E}$_{Opt}$ = $\Delta$\textit{E}$_{BGR}$ + $\Delta$\textit{E}$_{Bind}$ \begin{flushright}= \textit{E}$_{BGR}^{\epsilon_{2}=1}$($\epsilon_{2,Final}^{-1}$ - $\epsilon_{2,Initial}^{-1}$) + \textit{E}$_{Bind}^{\epsilon_{2}=1}$($\epsilon_{2,Final}^{-1.2}$ - $\epsilon_{2,Initial}^{-1.2}$).\end{flushright}

\begin{figure}
\begin{center}
\includegraphics[width=3.3in]{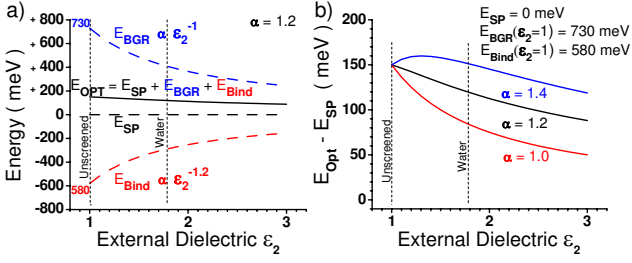}
\caption{(a) Scaling of particle interaction energies with external screening.  \textit{E}$_{BGR}$ is shown as a blue dashed line and scales with $\epsilon_{2}^{-1}$.  \textit{E}$_{BGR}^{\epsilon_{2}=1}$ is assumed to be 730 meV.  \textit{E}$_{Bind}$ is shown as a red dashed line and scales with $\epsilon_{2}^{-1.2}$.  \textit{E}$_{Bind}^{\epsilon_{2}=1}$ is assumed to be -580 meV.  The single particle energy, \textit{E}$_{SP}$, is taken to be zero.  The resulting behavior of \textit{E}$_{Opt}$ with screening is shown as a black line.  (b) The behavior of \textit{E}$_{Opt}$ (minus the constant \textit{E}$_{SP}$) as a function of $\alpha$.  Notice larger values of $\alpha$ lead to blue shifts of \textit{E}$_{Opt}$ with small screening.}
\label{fig3}
\end{center}
\end{figure}

Fig. \ref{fig3}a depicts the scaling behavior of the constituent particle interaction energies.  \textit{E}$_{BGR}^{\epsilon_{2}=1}$ is taken as 730 meV and \textit{E}$_{Bind}^{\epsilon_{2}=1}$ as 580 meV from Ref. \cite{Walsh}, where these are the interaction energies associated with the second valence band to second conduction band transition, E$_{22}$.  The values $\epsilon_{2}$ = 1 and 1.78 (water) are highlighted with vertical dashed lines.  It is important to note the relatively weak dependence of \textit{E}$_{Opt}$ with screening due to the opposite sings of \textit{E}$_{BGR}$ and \textit{E}$_{Bind}$, in accordance with the picture described in Refs. \cite{Ando,Kane1} and reported in the literature\cite{Lefebvre1,Moore}.  First order, single particle models, such as the nearest-neighbor tight-binding model, predict a constant ratio of \textit{E}$_{22}$ / \textit{E}$_{11}$ equal to 2.  Experiments found that ratio, on average, to be closer to a value of 1.7 \cite{Bachilo} and was dubbed the ``ratio problem''.  As explained in Ref. \cite{Kane2}, particle interaction energies resolve this discrepancy since \textit{E}$_{BGR}$ and \textit{E}$_{Bind}$ do not exactly cancel but lead to an overall blue shift of \textit{E}$_{Opt}$ compared to \textit{E}$_{SP}$, for each subband \textit{E}$_{ii}$.  That is, both the numerator and the denominator in the ratio \textit{E}$_{22}$ / \textit{E}$_{11}$ are slightly blue shifted and thus the ratio is decreased (ignoring chirality effects.)  Note, however, that the free particle band gap is significantly altered compared to the single particle value and can change dramatically upon perturbation of the environment.  This is, of course, a critical consideration with regard to CNT electronic device design and operation.  Fig. \ref{fig3}b shows the effect of the scaling exponent, $\alpha$, on the behavior of \textit{E}$_{Opt}$.  For the same \textit{E}$_{BGR}^{\epsilon_{2}=1}$ and \textit{E}$_{Bind}^{\epsilon_{2}=1}$, as in Fig. \ref{fig3}a, notice how larger values of $\alpha$ can actually lead to blue shifts in \textit{E}$_{Opt}$ in the small screening limit.\\

{\bf 3. Experiment and Results}\\

We have previously published experimental results interpreted using the above described model\cite{Walsh}. Resonant Raman spectroscopy, or RRS, was used to determine the optical transition energies, \textit{E}$_{22}$, for two single CNTs suspended across trenches (in order to remove substrate effects) as the dielectric environment was altered from dry N$_{2}$, to high humidity N$_{2}$, to water. The RRS experimental apparatus and technique is detailed in Ref. \cite{Yin1}.

\begin{figure}
\begin{center}
\includegraphics[width=3.3in]{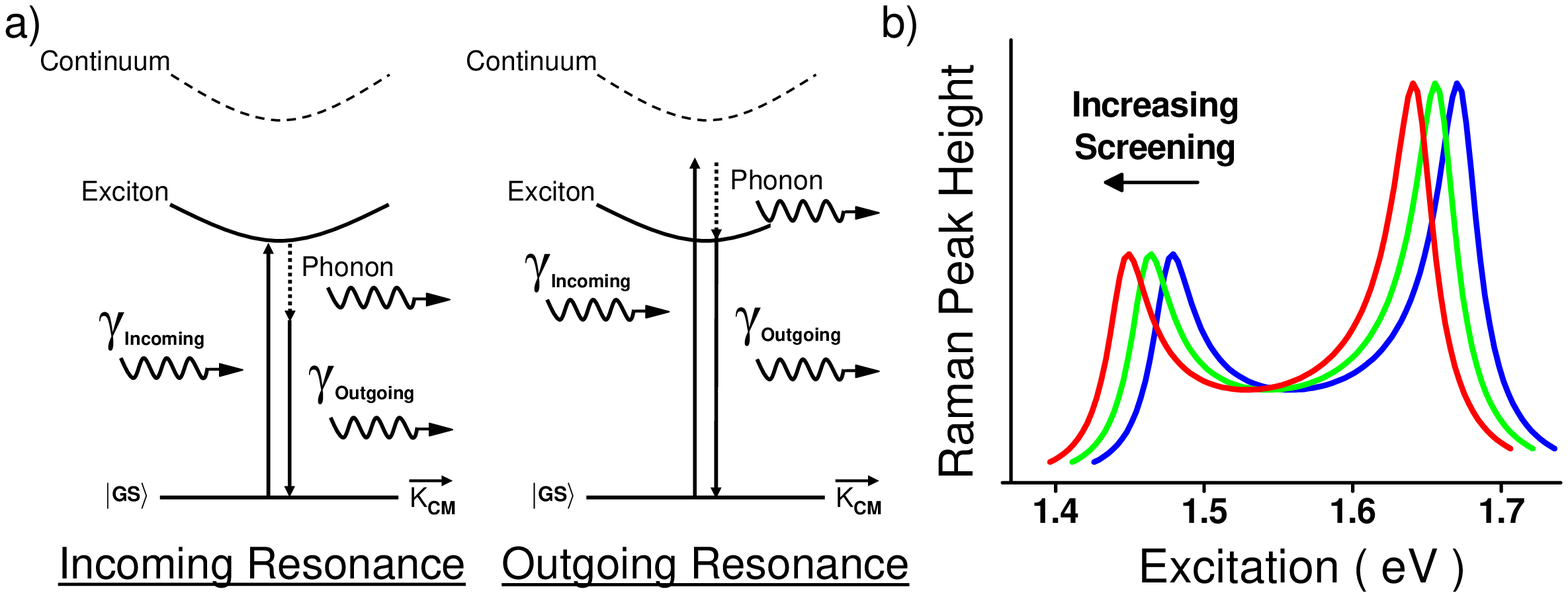}
\caption{(a) Energy level diagrams of the incoming and outgoing photon resonance conditions for the Stokes process.  (b) Cartoon of the dependence of the Raman excitation profile on external screening.}
\label{fig4}
\end{center}
\end{figure}

RRS has two significant advantages over photoluminescence (PL) spectroscopy in relation to this experiment.  First, PL spectroscopy is necessarily restricted to the lowest sub-band \textit{E}$_{11}$ whereas RRS can be used to probe the effect of screening on the exciton associated with any sub-band.  Further, most dielectric environments will tend to quench the PL signal almost entirely making the type of measurements made in this paper impossible using PL.  In RRS, plotting the Raman peak height for a given Raman active mode against laser excitation energy yields, in general, a curve with two peaks, known as the Raman excitation profile, or REP.  Figure \ref{fig4}a depicts the two different resonance conditions resulting in the two peaks in the REP.  Figure \ref{fig4}b shows a cartoon of the REP where the phonon energy is large enough that the two peaks are spectrally separated, as is the case with the Raman active G$_{+}$ vibrational mode.  The effect of screening is to red-shift the underlying optical transition energy and thus the REP red-shifts as well.  The REP can be fit using a one-phonon, exciton-mediated lineshape\cite{Vamivakas} for each dielectric condition and the shift in the optical transition energy determined.

Results showed a monotonic decrease of the optical transition energy with increasing external dielectric\cite{Walsh}, thus supporting a value of $\alpha$ $<$ 1.4 in accordance with Fig. \ref{fig3}b.  Using the model detailed here, \textit{E}$_{BGR}^{\epsilon_{2}=1}$ and \textit{E}$_{Bind}^{\epsilon_{2}=1}$ were found to be approximately 730 meV and 580 meV, respectively, at \textit{E}$_{22}$, for the (12,4) CNT.  That is, \textit{E}$_{BGR}^{\epsilon_{2}=1}$ was found to be greater than \textit{E}$_{Bind}^{\epsilon_{2}=1}$ by $\sim$ 150 meV in agreement with explanations of the ``ratio problem.''  Upon screening, these energies decreased significantly, of order 50\%.\\

{\bf 4. Conclusions}\\

We have derived the scaling relationship between exciton binding energy and the value of the external dieletric which is of practical importance when designing a variety of CNT based technologies.  The scaling exponent, $\alpha$ , is found to have a value of 1.2.  The relevant particle interaction energies are shown to decrease on order of 50\% upon screening by water.  The optical transition energy scales much more weakly with increasing external dielectric but could actually blue shift, in the small scale limit, with larger values of $\alpha$.\\

{\bf Acknowledgements}\\

The authors wish to thank An Slachmuylders for her help with corrections to Ref. \cite{Banyai}.  This work was supported by Air Force Office of Scientific Research under Grant No. MURI F-49620-03-1-0379, by NSF under Grant No. NIRT ECS-0210752, and by a Boston University SPRInG grant.


\begin{thebibliography}{25}
\expandafter\ifx\csname natexlab\endcsname\relax\def\natexlab#1{#1}\fi
\expandafter\ifx\csname bibnamefont\endcsname\relax
  \def\bibnamefont#1{#1}\fi
\expandafter\ifx\csname bibfnamefont\endcsname\relax
  \def\bibfnamefont#1{#1}\fi
\expandafter\ifx\csname citenamefont\endcsname\relax
  \def\citenamefont#1{#1}\fi
\expandafter\ifx\csname url\endcsname\relax
  \def\url#1{\texttt{#1}}\fi
\expandafter\ifx\csname urlprefix\endcsname\relax\def\urlprefix{URL }\fi
\providecommand{\bibinfo}[2]{#2}
\providecommand{\eprint}[2][]{\url{#2}}

\bibitem[{And()}]{Ando}
\bibinfo{note}{J. Ando, J. Phys. Soc. Jpn. 66 (1997) 207401.}

\bibitem[{Kan({\natexlab{a}})}]{Kane1}
\bibinfo{note}{C. L. Kane, E. J. Mele, Phys. Rev. Lett. 93 (2004) 197402.}

\bibitem[{Kan({\natexlab{b}})}]{Kane2}
\bibinfo{note}{C. L. Kane, E. J. Mele, Phys. Rev. Lett. 90 (2003) 207401.}

\bibitem[{Per()}]{Perebeinos}
\bibinfo{note}{V. Perebeinos, J. Tersoff, P. Avouris, Phys. Rev. Lett. 92
  (2004) 257402.}

\bibitem[{Spa()}]{Spataru}
\bibinfo{note}{C. D. Spataru, S. Ismail-Beigi, L. X. Benedict, S. G. Louie,
  Phys. Rev. Lett. 92 (2004) 077402.}

\bibitem[{Wan()}]{Wang2}
\bibinfo{note}{F. Wang, G. Dukovic, L. E. Brus, T. F. Heinz, Science 308 (2005)
  838.}

\bibitem[{Mau()}]{Maultzsch1}
\bibinfo{note}{J. Maultzsch, R. Pomraenke, S. Reich, E. Chang, D. Prezzi, A.
  Ruini, E. Molinari, M. S. Strano, C. Thomsen, C. Lienau, Phys. Rev. B 72
  (2005) 241402.}

\bibitem[{Duk()}]{Dukovic}
\bibinfo{note}{G. Dukovic, F. Wang, D. Song, M. Y. Sfeir, T. F. Heinz, L. E.
  Brus, Nano Lett. 5 (2005) 2314.}

\bibitem[{Cap()}]{Capaz3}
\bibinfo{note}{R. B. Capaz, C. D. Spataru, S. Ismail-Beigi, S. G. Louie, Phys.
  Rev. B 74 (2006) 121401.}

\bibitem[{Wal()}]{Walsh}
\bibinfo{note}{A. G. Walsh, A. N. Vamivakas, Y. Yin, S. B. Cronin, M. S. Unlu,
  B. B. Goldberg, A. K. Swan, Nano Lett. 7 (2007) 1485.}

\bibitem[{Lef()}]{Lefebvre1}
\bibinfo{note}{J. Lefebvre, J. M. Fraser, Y. Homma, P. Finnie, Appl. Phys. A.
  78 (2004) 1107.}

\bibitem[{Moo()}]{Moore}
\bibinfo{note}{V. C. Moore, M. S. Strano, E. H. Haroz, R. H. Hauge, R. E.
  Smalley, Nano Lett. 3 (2003) 1379.}

\bibitem[{Cho()}]{Choi}
\bibinfo{note}{J. H. Choi, M. S. Strano, Appl. Phys. Lett. 90 (2007) 223114.}

\bibitem[{Vam()}]{Vamivakas}
\bibinfo{note}{A. N. Vamivakas, A. G. Walsh, Y. Yin, M. S. Unlu, B. B.
  Goldberg, A. K. Swan, Phys. Rev. B 74 (2006) 205405.}

\bibitem[{Oga({\natexlab{a}})}]{Ogawa1}
\bibinfo{note}{T. Ogawa, T. Takagahara, Phys. Rev. B 44 (1991) 8138.}

\bibitem[{Ban()}]{Banyai}
\bibinfo{note}{L. Banyai, I. Galbraith, C. Ell, H. Haug, Phys. Rev. B 36 (1987)
  6099.}

\bibitem[{Ped({\natexlab{a}})}]{Pedersen2}
\bibinfo{note}{T. G. Pedersen, Phys. Rev. B 67 (2003) 113106.}

\bibitem[{Taf()}]{Taft}
\bibinfo{note}{E. A. Taft, H. R. Philipp, Phys. Rev. 138 (1965) A197.}

\bibitem[{Hay()}]{Hayashi}
\bibinfo{note}{H. Hayashi, N. Watanabe, Y. Udagawa, C.-C. Kao, P. Natl. Acad.
  Sci. U.S.A. 97 (2000) 6264.}

\bibitem[{Lou()}]{Loudon}
\bibinfo{note}{R. Loudon, Am. J. Phys. 27 (1959) 649.}

\bibitem[{Com()}]{Combescot}
\bibinfo{note}{M. Combescot, T. Guillet, Eur. Phys. J. B 34 (2004) 9.}

\bibitem[{Oga({\natexlab{b}})}]{Ogawa2}
\bibinfo{note}{T. Ogawa, T. Takagahara, Phys. Rev. B 43 (1991) 14325.}

\bibitem[{Ped({\natexlab{b}})}]{Pedersen1}
\bibinfo{note}{T. G. Pedersen, Phys. Rev. B 67 (2003) 073401.}

\bibitem[{Bac()}]{Bachilo}
\bibinfo{note}{S. M. Bachilo, M. S. Strano, C. Kittrell, R. H. Hauge, R. E.
  Smalley, R. B. Weisman, Science 298 (2002) 2361.}

\bibitem[{Yin()}]{Yin1}
\bibinfo{note}{Y. Yin, A. G. Walsh, A. N. Vamivakas, S. B. Cronin, A. M.
  Stolyarov, M. Tinkham, W. Bacsa, M. S. Unlu, B. B. Goldberg, A. K. Swan, IEEE
  J. Sel. Top. Quantum Electron. 12 (2006) 1083.}

\end{thebibliography}
\end{document}